\documentclass[11pt,iaus]{amsart}
\usepackage{geometry}                
\usepackage{graphicx}
\usepackage{amssymb}
\usepackage{epstopdf}

\title{Cosmic Explosions (Optical Transients)}
\author{S. R. Kulkarni\\
Caltech Optical Observatories\\
Pasadena, California 91125, USA}

\begin{document}
\maketitle

One of the principal motivations of wide-field and synoptic surveys
is the search for and study of transients. By transients I mean
those sources which arise from the background, are detected, and
then fade away to oblivion.  Transients in distant galaxies need
to be sufficiently bright so as to be detectable and in almost all
cases these transients are catastrophic events, marking the deaths
of stars. Exemplars include supernovae and gamma-ray bursts.  In
our own Galaxy, the transients are, in almost all cases, cataclysmic
rather than catastrophic, e.g. flares from M dwarfs, novae of all
sorts (dwarf novae, recurrent novae, classical novae, X-ray novae)
and instabilities in the surface layers (S Dor, eta Carina).  In
the nearby Universe (say out to the Virgo cluster) we have sufficient
sensitivity to see classical novae.

This paper is an extended summary of the talk I gave at IAU Symposium
\textit{New Horizons in Time Domain Astronomy} (Oxford, 2011).\footnote{The
paper is focused on the search for and study of transients. The
needs for this field are quite different from that of variable
stars.  The reader should bear this distinction in mind when reading
this paper.} I first review the history of transients (which is
intimately related to the advent of wide-field telescopic imaging;
\S\ref{sec:HistoryPhaseSpace}). In \S\ref{sec:EraSynopticSurveys}
I summarize wide field imaging projects. The motivations that led
to the design of the Palomar Transient Factory (PTF) followed by a
summary of  the astronomical returns can be found in \S\ref{sec:PTF}.
In \S\ref{sec:LessonsLearnt} I review the lessons learnt from PTF.
I conclude that, during this decade, optical transient searches
will continue to flourish and may even accelerate as surveys at
other wavelengths -- notably radio, UV and X-ray --  come on line.
As a result, I venture to suggest that specialized searches for
transients will continue -- even into the LSST era.  I end the
article by discussing the importance of follow-up telescopes for
transient object studies -- a topical issue given that in the US
the Portfolio Review is under away (\S\ref{sec:BeetlesBehemoths}).

\section{History: Wide Field Imaging \&\ Phase Space}
\label{sec:HistoryPhaseSpace}

One could say that large surveys of the sky for star positions,
stellar photometry and stellar classification (via low resolution
spectroscopy)  essentially define the start of the modern era of
astronomy. The early returns resulting from the discovery and study
of variable stars were stunning. RR Lyrae and Cepheid variables
proved to be rather precise yardsticks and astronomers came to
appreciate the physical scale of our Galaxy and eventually the
physical scale of the Local Universe.

\begin{figure}[hbtp] 
   \centering
   \includegraphics[width=3in]{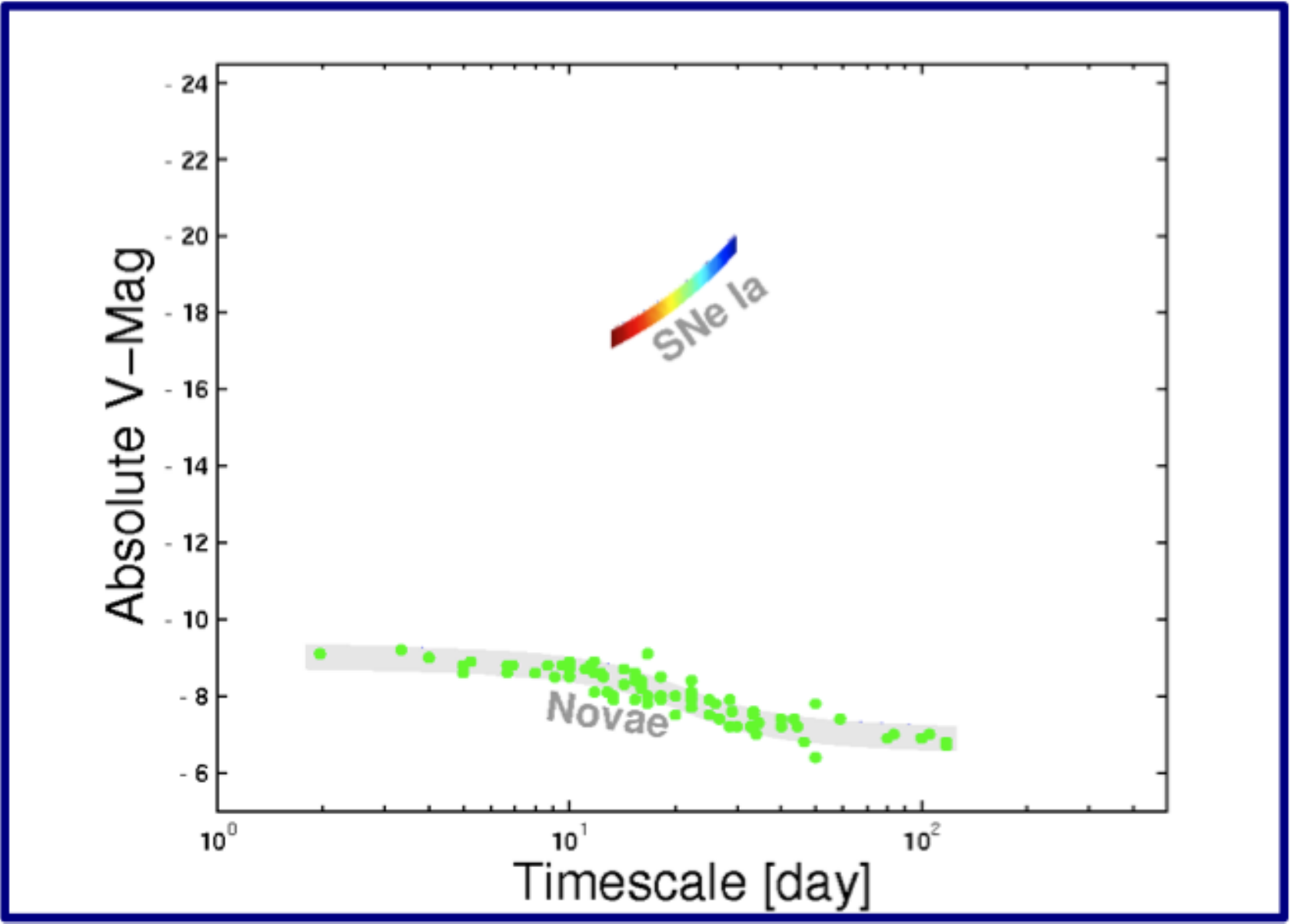}
   \caption{\small Phase space of cosmic   explosions in the Zwicky
   era: novae and Ia supernovae.  An explosion has several basic
   parameters: energy of explosion, mass of ejecta, velocity of
   ejecta, rise time of explosion, peak luminosity and  decay time
   of explosion. Peak luminosity and decay timescale are easily
   measured and therefore constitute the principal axes of the phase
   space of transients.  The horizontal axis is the decay timescale
   (1\,day to 1\,year)  and the y-axis is the peak luminosity but
   shown as absolute magnitude in the V-band.  Type II explosions
   have not been shown because the explosion physics is masked by
   the envelope. Notice the wide gap between novae and super-novae.}
   \label{fig:SimpleTauMv}
\end{figure}

All transients were initially classified as ``nova stella'' or new
stars and the abbreviation novae came to be specifically applied
to essentially classical novae.  Classical novae were sought for
their purported use in determining the distance to the Andromeda
galaxy (fast novae are, on the average, brighter than slow ones).
Telescopes on  Mt. Wilson were pressed into M31 novae observations
by E. Hubble and others.

The modern era of transients with controlled cadence and a physics-based
enquiry began with F. Zwicky and W. Baade. Recognizing the importance
of the then newly invented ``Schmidt'' type wide-field
telescope\footnote{Alerted to Zwicky by W. Baade who knew the
inventor Bernhard Schmidt.}, Zwicky obtained funds from a wealthy
family in Pasadena and had an 18-inch telescope  using the Schmidt
camera design constructed.  The ``P18''  was the first telescope
on the Palomar mountain\footnote{The telescope still exists and
will soon be moved to the Palomar Museum; the P18 dome now is our
aeronomy center and houses a polar telescope for seeing monitoring.}.

The first major result was the recognition of two distinct families:
classical novae and ``super-novae'' (Baade \& Zwicky 1934); see
Figure~\ref{fig:SimpleTauMv}.  In their very next paper the authors
made the bold conjecture that supernovae mark the transmutation of
an aging star into a neutron star, a most compact object (which
itself was a novel hypothesis first proposed by L. Landau in 1931).
The resulting enormous release of gravitational binding energy would
accelerate some particles to relativistic velocities or cosmic rays.
Next, thanks to the systematic survey carried out by F. Zwicky,
families of supernovae were recognized.\footnote{Type I and Type
II have certainly survived the passage of time. It would be interesting
to revisit types III, IV and V that Zwicky had proposed.}

The success of P18 motivated Zwicky to seek a larger Schmidt-type
telescope and this led to the 48-inch telescope\footnote{Now called
the Samuel Oschin telescope after a benefactor whose gift allowed
the telescope to be rejuvenated.} (P48). This telescope saw first
light at about the same time as the Palomar 200-inch (circa 1948).
P48 undertook two ambitious and comprehensive Northern hemisphere
sky surveys (POSS-I and POSS-2).  These surveys had a fundamental
impact on astronomy and inspired subsequent all sky surveys.
Following the POSS program the telescope was modernized (Pravdo et
al. 2000) and the photographic plates were replaced by CCDs with
ever increasing sophistication: NEAT (Pravdo et al. 2000); QUEST
(Baltay et al. 2007); upgraded CFH12K (Rahmer et al. 2008).

\section{The Era of (Optical) Synoptic Surveys}
\label{sec:EraSynopticSurveys}

It is now clear that we are well into the era of synoptic and/or
wide field astronomy.  In the 1-m to 2-m category we have the
Catalina Sky Survey (CSS), PTF, Pan-STARSS-1 (PS-1), La~Silla QUEST
and SkyMapper. In the 2-m to 4-m category we have MegaCam/CFHT, the
VLT Survey Telescope, Dark Energy Camera/Blanco [expected commissioning:
2012] and the One Degree Imager/WIYN (under construction). Finally
in the behemoth category we have Suprime-Cam and soon Hyper-Suprime-Cam
[2012] on the Subaru 8.3-m telescope.  The Large Synoptic Survey
Telescope (LSST), expected by the end of the decade, is also an
8.3-m telescope,  but equipped with an imager with a field-of-view
about five times larger than that of Hyper Suprime-Cam.

\section{The Palomar Transient Factory (PTF)}
\label{sec:PTF}

PTF consists of two dedicated telescopes (see Law et al. 2009): the
P48 equipped with a re-engineered CFH12K (Cuillandre et al. 2001)
96-Megapixel CCD mosaic (Rahmer et al. 2008) and a field-of-view
of 7.2 square degrees acting as the Discovery Engine and the Palomar
60-inch (P60) equipped with a 4-Megapixel CCD camera acting as the
Photometric Engine.\footnote{PTF is a collaboration of the following
entities: California Institute of Technology, Lawrence Berkeley
Laboratory (LBL), Weizmann Institute of Sciences, Las Cumbres
Observatory Global Telescope, Columbia University, Oxford University,
Infrared Processing \&\ Analysis Center (IPAC) \&\ UC Berkeley.
The responsibilities are as follows: image subtraction pipeline
(LBL), photometric pipeline (IPAC), classification (UC Berkeley),
P60 robotization and operations (D. Fox, B. Cenko \&\ M. Kasliwal),
P48 sequencer (E. Ofek), PTF Marshal (R. Quimby) and spectroscopic
reduction \&\ archive (A. Gal-Yam).}

\begin{figure}[htbp]
   \centering
   \includegraphics[width=3in]{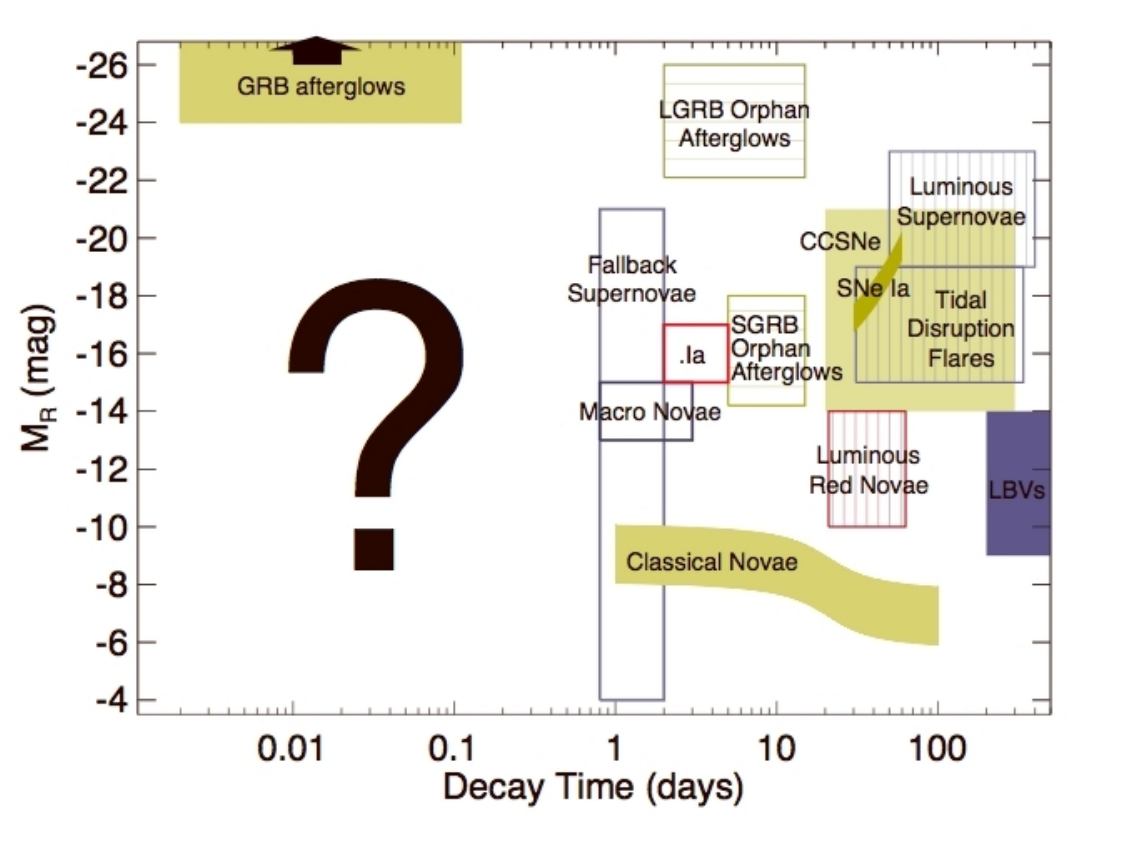}
   \caption{\small Theoretical and physically plausible candidates
   are marked in the explosive transient phase space. The original
   figure is from Rau et al. (2009). The updated figure (to show
   the unexplored sub-day phase space) is from the LSST Science
   Book (v2.0). Shock breakout is the one assured phenomenon on the
   sub-day timescales. Exotica include dirty fireballs, newly minted
   mini-blazars and orphan afterglows. With ZTF we aim to probe the
   sub-day phase space (see \S\ref{sec:BeetlesBehemoths}).  }
\label{fig:TransientPhaseSpace} \end{figure}

PTF was motivated by two considerations (Rau et al. 2009).  First,
is the exploration of transients in the sky. With reference to
Figure~\ref{fig:SimpleTauMv} our goal was to find objects in the
nova-supernova gap (for which a number of physically motivated
scenarios exist). A second motivation is the great promise of
entirely new areas of astronomy: 
{\small 
\begin{enumerate} 
\item High Energy Cosmic Rays.  
\item High Energy Neutrinos.  
\item Highest Energy Photons.  
\item Gravitational Wave Astronomy.  
\end{enumerate}
} 
The sources of interest to these facilities are connected to
spectacular explosions. However, the horizon (radius of detectability),
either for reasons of optical depth (GZK cutoff; $\gamma\gamma\rightarrow
{\rm e}^{\pm}$) or sensitivity, is limited to the Local Universe
(say, distance $\lesssim 100$\,Mpc). Unfortunately, these facilities
provide relatively poor localization. The study of explosions in
the Local Universe  is thus critical for two reasons: (1) sifting
through the torrent of false positives (because the expected rates
of sources of interest is a tiny fraction of the known transients)
and (2) improving the localization via low energy observations
(which usually means optical).  In Figure~\ref{fig:TransientPhaseSpace}
we display the phase space informed by theoretical considerations
and speculations. Based on the history of our subject we should not
be surprised to find, say a decade from now, that we were not
sufficiently imaginative.  

The clarity afforded by our singular focus -- namely the exploration
of the transient optical sky -- allowed us to optimize PTF for
transient studies. Specifically, we undertake the search for
transients in a single band (R-band during most of the month and
$g$ band during the darkest period). As a result our target throughput
is five times more relative to multi-color surveys (e.g. PS-1,
SkyMapper).

Given the ease with which transients (of all sorts) can be detected, 
in most instances, the transient without any additional information
for classification does not represent a useful, let alone a meaningful, advance.  
It is useful here to make the clear detection between 
\textit{detection}\footnote{ By which I mean that a transient has
been identified with a reliable degree of certainty.}
and \textit{discovery}.\footnote{By which I mean that the
astronomer has a useful idea of the nature of the transient. At the
very minimum we should know if the source is Galactic or extra-galactic.
At the next level, it would be useful to have the first level of
sub-typing (eg. flare star/DN/CV; Ia/Ibc/II SN). This knowledge is essential
given the very large fog of foreground (M dwarf flares, dwarf noave)
and background transients (routine supernovae at a late phase, burps from an AGN).} Thus the burden for
discovery is considerable since for most transients this would require
spectroscopy.  
At the final level is
a clear pigeon holing of the transient (classification).
The importance 
of this point was re-iterated, even more forcefully, in the concluding
talk (Bloom  2011). It is frustrating to hear some astronomers, especially at
august meeting such as this, to claim a discovery merely on the
basis that they had observed the transient earlier than others.

Recognizing the above issue we
adopted a ``No Transient Left Behind'' strategy.  Three-color
photometry on P60 allows for crude classification. Follow up up
with low resolution spectroscopy on a bevy of larger telescopes
(Palomar 200-inch, KPNO 4-m, WHT 4.2-m and the Lick 3-m)\footnote{Even
so, as in real life, two thirds of the transients are unclassified
and left behind.}. As a result we have amassed a set of nearly 1500
\textit{spectroscopically classified} supernovae of which a good
fraction were detected prior to maximum.

\begin{figure}[htbp] 
   \centering \includegraphics[width=3.6in]{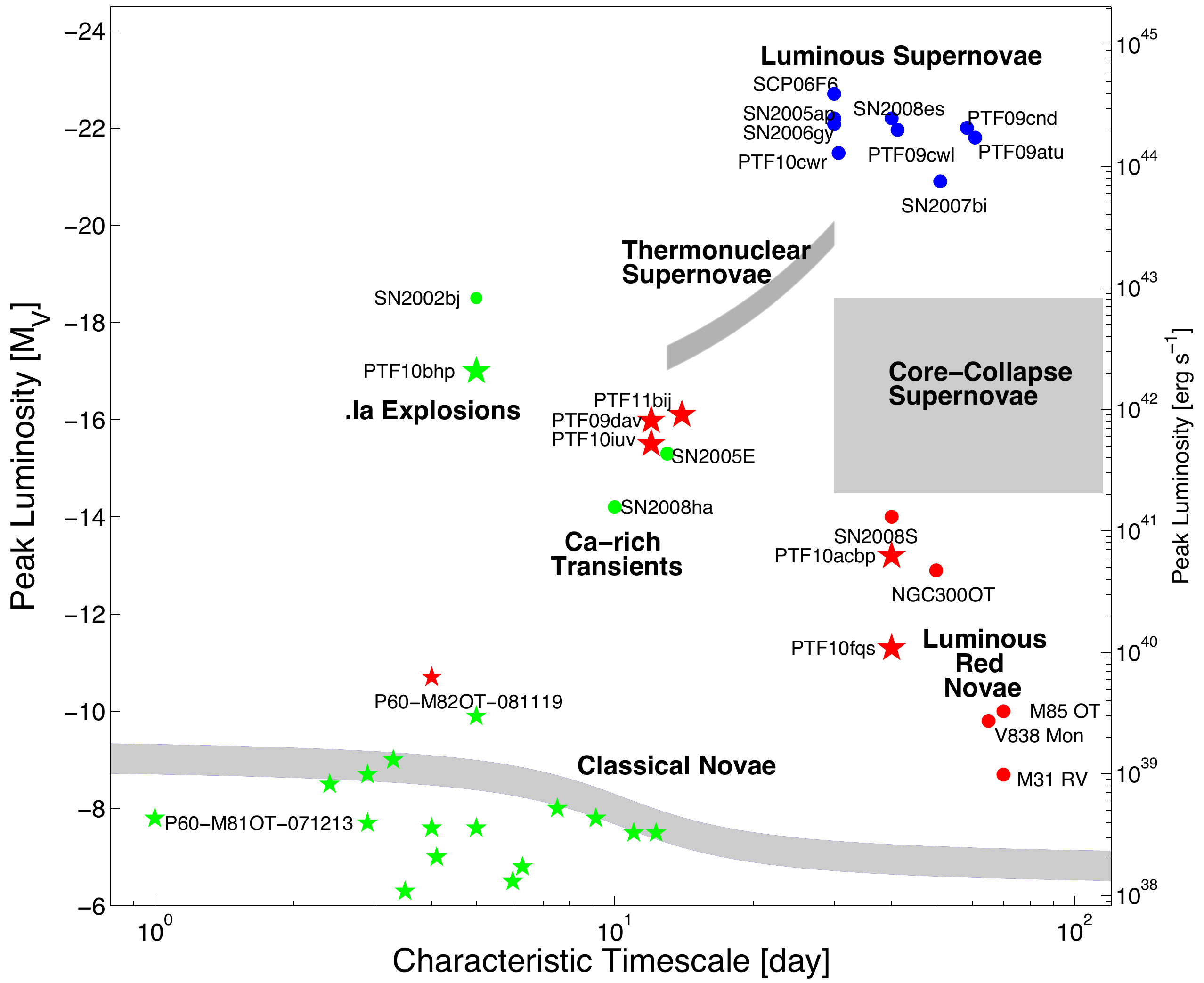}
   \caption{\small An update of Figure~\ref{fig:SimpleTauMv} and
   \ref{fig:TransientPhaseSpace}
    with new classes and sub-classes of non-relativistic transients.
    Notice the emerging class of Calcium-rich halo transients, .Ia
    supernovae and two types of Luminous Red Novae (events in the bulges
    of M85 and V838 Mon; the others are in spiral arms).
     The color of the symbol is that at maximum light.
     Apparently the new data show that novae do not obey the classical
     ``Maximum Magnitude Rate of Decay''  relation (see Kasliwal 2011).
   } \label{fig:ObservedPhaseSpace}
\end{figure}

Given that follow-up is at premium having a small sample of transients
with desired or well-understood selection criteria is more valuable
than a large sample of transients with a potpourri of properties.
Thus choice of pointings and cadence control are critical. We have
scoured around the sky to select PTF pointings with large local
($d\lesssim 200\,$Mpc) over densities.  The  nearly one hundred
selected pointings contain $\times 4$  more light than randomly
chosen pointings (Kasliwal, 2011).  Cadence control is even more
important. For instance the logistics of obtaining UV spectra of
nearby Ia supernovae (important for calibration of Ia cosmology)
require that the supernovae be identified 10 to 14 days prior to
peak.  In 2010 we focused on finding such supernovae and we now
have three dozen HST UV spectra of nearby Ia -- an order of magnitude
in the sample size (Cooke et al. 2011).

Another goal  is to decrease the  latency between detection and
discovery. During 2011 we
made efforts to decrease the latency and were (very luckily) rewarded
with the discovery of a Ia supernova in the very local Universe,
PTF11kly in Messier~10,  just 11 hours after the explosion (Nugent
et al.  2011).  The proximity and near natal discovery allowed us
and other astronomers to shed fundamental new light on the progenitor
of a Ia supernova.

With 38 refereed
publications\footnote{\texttt{http://www.astro.caltech.edu/ptf}.
PTF began routine operations summer of 2009. Half of the Fall season
was lost to the Great Station Fire.} (a mean rate of 1 publication
per month) PTF has been productive.  Five PhD theses\footnote{ M.
Kasliwal -- Transients in the Local Universe; I. Arcavi -- Demographics
of Core-collapse Explosions; D. Levitan -- AM CVn stars; S. Ben-Ami
-- Early Emission from SN; A. Waszczak -- Small bodies in the Solar
System} and nearly a dozen of postdocs have been and are being
supported.  In addition to the two results summarized above, key
discoveries and findings include the clear identification of a class
of luminous supernovae whose spectra show no hydrogen (Quimby et
al. 2011), the gradual ``coloring of the phase space''
(Figure~\ref{fig:ObservedPhaseSpace}), the demographics of core
collapse supernovae (Arcavi et al. 2010) and an apparently exotic
nuclear event (PTF10iya; Cenko et al.  2011).

\section{Lessons Learnt}
\label{sec:LessonsLearnt}

First and foremost,  a clear vision for the project must be
articulated.  In this regard, Figure~\ref{fig:TransientPhaseSpace}
was an essential exercise\footnote{In this respect we followed
Zwicky's morphological approach to problem solving which stresses
the importance of a sensible (consistent with physics) exploration
of phase space and not merely an aspiration based approach.} in
motivating PTF. As discussed earlier (\S\ref{sec:PTF}) detection
is merely the first (and easy) step.\footnote{In fact, I predict, that increasingly the same transient will be detected
in parallel by more than on-going survey. Surveys with high cadence control will
have an advantage over multi-purpose surveys.}  Discovery requires much harder work and 
in almost all cases follow up spectroscopy (access to follow up telescopes) and a deep
knowledge of astronomy (expert knowledge) is needed. These two demands
can easily exceed that needed for the search 
itself.

Next, cadence control is absolutely essential to produce quality
transients worthy of follow-up.  Fourth, a horizontal structure
with essentially independent key project teams allows for motivated
teams to undertake efficient and rapid follow-up.  A corollary is
that strong scientific leadership is essential to resolve overlapping
interests. Fifth -- a truism -- software pipelines have to be fully
functional before the searches begin.  Sixth,  as illustrated by
the case of PTF11kly (\S\ref{sec:PTF}), the collaboration should
be flexible enough to take advantage of new findings and/or organic
advances in the field.

We plan to strictly implement these lessons in the next phase of
PTF (``iPTF'') which will run for 2013 \&\ 2014. For this phase we
are on course to completing a high throughput ultra-low resolution
($\lambda/\Delta\lambda\approx 100$) IFU spectrograph (SED
Machine)\footnote{\texttt{http://sites.google.com/site/nickkonidaris/sed-machine}.
The SEDM will replace the imaging photometer.} and optimized for
spectral classification.

\section{Future: Beetles \&\ Behemoths}
\label{sec:BeetlesBehemoths}

One may reasonably ask: is there a need for more synoptic surveys
since even a modest aperture synoptic survey can generate more data
than can be digested? This is a meritorious question. My reply is
that there is a \textit{need} for highly focused projects. I list
three case studies.  CSS, a survey based on 1-m telescope equipped
with a routine CCD detector, is remarkable for its harvest of NEOs
and pinpointing (with considerable heads-up) the fiery entry of
2008TC3 (Boattini et al.  2009).  The case for PTF with its laser
like focus on transients can be found in \S\ref{sec:PTF}. CFHT SN
Legacy Survey (CFHT; Conley et al. 2011) has made deep contributions
to Ia cosmology. The Medium Deep Survey of PS-1 has proved to be
adept at  identifying luminous supernovae (Chomiuk et al. 2011).
The nightly cadenced SDSS Stripe~82 project has made contributions
across board -- SN cosmology, AGN variability, stellar variability,
tidal disruption events (e.g. van Velzen et al. 2011).  It appears
that focused projects have the ability to trump general-purpose
larger facilities which are subject to competing cadence demands.

The second reason that I remain bullish of modest aperture searches
is that the earnest exploration of the phase space has just begun.
The phase space with decay time of less than a day (for which there
are several plausible scenarios) and the entire physics of the rise
time of explosive transients  is essentially wide open.

Finally, currently, the transient game is  dominated by optical
synoptic facilities and projects. The next to join this club\footnote{The
optical synoptic revolution is a result of Moore's law for computing
and optical sensors. The radio revolution is being driven by
exponential advances in computing, RF technology, advances in
interferometric imaging algorithms and LNSD architecture.} would
be radio facilities at both meter and decimeter wavelengths (EVLA,
LOFAR, MWA, upgraded GMRT, APERTIF, MeerKAT). With some luck synoptic
space based projects may also happen during this decade (e.g.
LIMSAT\footnote{ This is a cluster of 12-cm aperture telescopes
with a total of 1,000 square degrees and promoted by an ad hoc group
of astronomers from Israel, Caltech, India and Canada.} in the UV,
rejuvenation and repurposing of WISE for thermal IR searches and a
Lobster-type mission in the X-rays).  One can easily imagine joint
studies with optical facilities.

Projects such as PTF, CSS and PS-1 demonstrate that there is ample
scope for 1-m to 2-m class surveys to continue well into this decade.
The magnitude limit of 21 is ideally suited to classification
spectroscopy. Going fainter is only an advantage if one has the
ability to discriminate between the ``unknown unknowns'' against
the dense  fog of known transients or if there is a compelling
reason to do so (e.g. using luminous supernovae as tracers of star
formation).  Recall that follow-up at the 22-mag requires 6 times
as much as follow-up time  as a 21-mag event.

Currently, in the US there is a debate on the future of existing
facilities. In Astronomy, discoveries keep the field exciting and
large telescopes (especially spectroscopy) and theory provide the
physical understanding. The reasons to build increasingly larger
telescopes and the motivation for great facilities has and will
remain strong. However, discoveries primarily result from surveys
(e.g. discovery of very high redshift quasars from SDSS and UKIRT)
and/or concerted efforts on modest-size telescopes (e.g. planets
around normal stars; the discovery of the first brown dwarf).

Hopefully the reader is by now convinced that the discovery potential
for the sub-field of transients has been and will continue to be
large. As such one should let many ideas bloom.  The cost of searches
with modest-size telescope is affordable.  For instance, the total
cost of PTF (capital and 4-year operating costs) is under \$3M.
Repurposing the present collection of 2-m to 4-m existing telescopes
to support (mainly light curves and spectroscopy) the on-going
surveys is a cost effective way to continue doing cutting-edge
research. The current  imbalance of big/expensive/facility over
small/focused projects is neither cost effective nor strategic and,
as is now being realized, financially not sustainable.

So enthused I am with the prospects and promises of focused transient
searches with assured  follow-up  that, along with my colleagues,
I am now proposing a second generation of PTF. We propose to fully
populate the focal plane of P48 (40 square degree) and equip two
other telescopes with the SED Machine (for rapid spectral classification;
\S\ref{sec:LessonsLearnt}) and a robotic laser Adaptive Optics
system' system\footnote{\texttt{http://www.astro.caltech.edu/Robo-AO}}
for rapid photometry in crowded host galaxy fields. The primary
focus of this three-telescope \textit{facility} is to probe the
sub-day phase space (see Figure~\ref{fig:TransientPhaseSpace}).  I
propose to name this as the Zwicky Transient Facility (ZTF) after
the founder of our field.  First light is expected in 2015.  Should
interesting sub-minute bursts be discovered then we will  upgrade
ZTF with CMOS.  Hopefully ZTF, a beetle, will prove to be productive
even as general purpose behemoths come on line.

\bigskip\bigskip
\noindent{\it Acknowledgements:} {\small I would like to acknowledge
National Astronomy Observatory of Japan and the Japan Society for
Promotion of Science for hosting my sabbatical stay in Japan during
which this article was completed.

The hard work and creativity of the members of PTF have made the
project productive. I would like to specially acknowledge the
following colleagues: L. Bildsten, J. Bloom, S. B. Cenko, R. Dekany,
A. Gal-Yam, M. Kasliwal, R.  Laher, N. Law,  P. Nugent, E. Ofek,
R. Quimby, R. Smith \&\ J. Surace. The excellent staff of the Caltech
Optical Observatories made it possible for re-engineering of CFH12K
and refurbishment of P48.  The low downtime, despite the age of P48
and P60, is a testament to the amazing crew at Palomar mountain.

CHF12K gave PTF a great start. I truly appreciate the excellent
engineering of the builders of CFH12K and the generosity of C.
Veillet and the CFHT Corporation. I would like to thank D.  Frail,
G. Helou \&\ T. Prince for many discussions and help.  Finally, I
would like to thank W. Rosing whose initial  investment in PTF
allowed me to garner funds from other parties and thereby complete
the project in record time. }

\twocolumn
\bigskip
{\small

\noindent
Arcavi, I. et al. (2010), \textit{ApJ} 721, 777

\noindent
W. Baade \&\ F. Zwicky (1934), \textit{PNAS} 20, 254

\noindent
Baltay, C. et al. (2007), \textit{PASP}, 119, 1278 

\noindent
Boattini, A. et al. (2009), \textit{DPS} 41, \#9.02

\noindent
Bloom, J. et al. (2011), \textit{This Proceedings}

\noindent
Cenko, S. B. et al. (2011), arXiv1103.0779

\noindent
Chomiuk, L. et al. (2011), \textit{ApJ} 743, 114

\noindent
Cooke,  J. et al. (2011), \textit{ApJ} 727, L35

\noindent
Conley, A. et al. (2011), {\it ApJS} 192, 1

\noindent
Cuillandre, J-C., Starr, B., Isani, S. \&\ Luppino, G. (2001),
\textit{ExA} 11, 223

\noindent
Kasliwal, M. M. (2011), PhD Thesis, Caltech

\noindent
Law, N. et al. (2009), \textit{PASP} 121, 1395

\noindent
Nugent , P. et al. (2011), \textit{Nature} 480, 344

\noindent
Pravdo, S. et al. (2000), \textit{DPS} 32, 1640

\noindent
Quimby, R. et al. (2011), \textit{Nature} 474, 487

\noindent
Rau. A. et al. (2009), \textit{PASP} 121, 1334

\noindent
Rahmer, G. et al. (2008), \textit{SPIE}, 7014, 70144Y

\noindent
van Valzen, S. et al. (2011), \textit{ApJ} 741, 73
}
\end{document}